\documentclass[referee]{aa}
\usepackage{natbib,graphicx}

\newcommand{\be}{\begin{equation}}
\newcommand{\ee}{\end{equation}}

\begin{document}

\date{\today}

\title{Polarized dust emission \\
of magnetized molecular cloud cores}

\author{Jos\'e Gon\c{c}alves\inst{1,2}, Daniele Galli\inst{2}, 
\and Malcolm Walmsley\inst{2}}

\institute{Centro de Astronomia e Astrof\'{\i}sica da Universidade 
de Lisboa, Tapada da Ajuda, 1349-018 Lisboa, Portugal
\and INAF-Osservatorio Astrofisico di Arcetri, Largo E. Fermi 5, 
I-50125 Firenze, Italy}

\offprints{J.~Gon\c{c}alves, \\
\email{goncalve@arcetri.astro.it}}

\date{Received / Accepted}

\authorrunning{Gon\c calves, Galli, Walmsley}
\titlerunning{Polarization in cloud cores}

\abstract{We compute polarization maps for molecular cloud cores modeled
as magnetized singular isothermal toroids, under the assumption that
the emitting dust grains are aspherical and aligned with the
large-scale magnetic field. We show that, depending on the inclination
of the toroid with the line-of-sight, the bending of the magnetic field
lines resulting from the need to counteract the inward pull of gravity
naturally produces a depolarization effect toward the centre of the
map. We compute the decrease of polarization degree with increasing
intensity for different viewing angles and frequencies, and we show
that an outward increasing temperature gradient, as expected in
starless cores heated by the external radiation field, enhances the
decrease of polarization.  We compare our results with recent
observations, and we conclude that this geometrical effect, together
with other mechanisms of depolarization, may significantly contribute
to the decrease of polarization degrees with intensity observed in the
majority of molecular cloud cores. Finally, we consider the dependence
of the polarization degree on the dust temperature gradient predicted for externally
heated clouds, and we briefly comment on the limits of the
Chandrasekhar-Fermi formula to estimate the magnetic field strength in
molecular cloud cores.
}

\maketitle

\section{Introduction}

Mapping the polarization of the thermal emission of dust at millimitre
or submillimetre wavelengths (usually $\lambda=850$~$\mu$m or 1.3~mm)
is the principal means of probing the magnetic field geometry in molecular
cloud cores. A frequent characteristic of these observations
is the decrease of polarization degree $p$ as a function of the total
observed intensity $I$.  Usually $p$ decreases with increasing intensity
$I$ with a power-law behavior, from a maximum value of $\sim 15$\% to
about the observable limit of $\sim 1$\%.  This depolarization effect
(sometimes referred to as ``polarization hole'', or ``polarization limb
brightening'') has been observed in many dense cores and filamentary
clouds.  Recent examples include the OMC-3 region of the Orion A
filamentary molecular cloud where $p\propto I^{-0.7}$ (Matthews \&
Wilson~2000), several dense cores in the dark cloud Barnard~1, where
$p\propto I^{-0.8}$ (Matthews \& Wilson~2002), Bok globules mapped by
Henning et al.~(2001) where $p\propto I^{-0.6}$, and dense cores mapped
interferometrically by Lai et al.~(2002), where $p\propto I^{-0.8}$.
In some cases, the decrease of polarization with intensity is quite
steep: for example, Crutcher et al.~(2004) find $p\propto I^{-1.2}$
in the dark cloud L183, implying that not only the polarization degree
but also the polarized intensity $I_p=pI$ decreases toward the centre of
this cloud. A similar situation is apparently found also in the starless
core L1544 (Ward-Thompson et al.~2000). 

\begin{figure}[ht]
\resizebox{10cm}{!}{\includegraphics{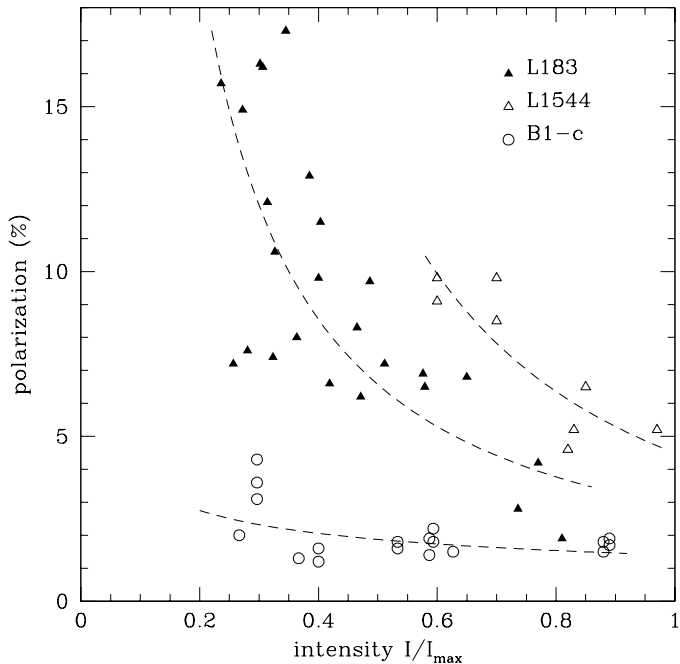}}
\caption{Distribution of percentage of polarization $p$ vs. intensity
$I/I_{\rm max}$ at $\lambda=850$~$\mu$m, normalized to its maximum value,
for the starless cores L183 ({\it filled triangles}, from Crutcher et
al.~2004), L1544 ({\it empty triangles}, from Ward-Thompson et al.~2000),
and the B1-c core ({\it circles}, from Matthews \& Wilson~2002). The
{\it dashed lines} are power-law fits, with slopes $-1.6$ for L1544,
$-1.2$ for L183, and $-0.4$ for B1-c.}
\label{observations}
\end{figure}

The dark cloud Barnard 1 is an excellent example of the type of object
of interest to us. This cloud, mapped with SCUBA by Matthews \&
Wilson~(2002), has dimensions of roughly $0.2 \times 0.4$~pc, densities
of a few times $10^4$~cm$^{-3}$, and temperatures of order $10$~K.
These appear unfavorable characteristics for aligning grains (see e.g.
Lazarian et al.~1997). Nevertheless, B1 exhibits extremely uniform
polarization at a level of around 5\% and with an extremely uniform
position angle. This last point in our opinion is a strong argument in
favor of an essentially uniform and likely dominant magnetic field over
most of the core. However, there are exceptions to the rule in the
shape of four high density (of order $10^6$~cm$^{-3}$) prestellar or
protostellar ``inclusions'' of dimensions $\sim 0.03$~pc (B1-a,b,c,d)
apparently embedded within the region of the ``general core'' referred
to above.  These extremely high density cores also surprisingly show
evidence for polarization, albeit at a lower level and with different
position angles than the general core.  In particular, in the core
B1-c the polarization degree is very low and almost uniform over the whole observed
intensity range (Matthews \& Wilson~2002) \footnote{ We note however
parenthetically that the observations of polarization at high densities
($n\sim 10^6$~cm$^{-3}$) where radiation fields can hardly be of
importance represent a serious challenge to grain alignment theorists.
In fact, recent work shows that radiative torques due to starlight are
required to drive grains to the suprathermal rotation rates necessary
to minimize the disaligning effect of random collisions (Draine \&
Weingartner~1996, 1997).}.  For illustration purposes, the $p$--$I$
relations measured in three well studied cores (L183, L1544, and B1-c)
are collected in Fig.~\ref{observations}, together with power-law fits
of the original data.

The observed decrease of polarization with increasing intensity has been
attributed to changes in the dust grain optical properties or shapes
in the cold cores interiors due to grain-growth (Vrba et al.~1993,
or Hildebrand et al.~1999): not only are bigger grains more difficult
to align than smaller grains, but also the agglomeration process
may make grains more spherical, thus further reducing the alignment
efficiency. 

All these effects are difficult to quantify. A larger density tends
in general to dealign grains (Lazarian et al.~1997), but the enhanced
magnetic field strength should have the opposite effect.  On the basis of
the Davis \& Greenstein~(1951) alignment model by paramagnetic relaxation,
Vrba et al.~(1981) showed that the ratio $p/A_V$ (equivalent to $p/I$
at millimetre wavelengths) should scale as $B^2 a^{-1} n^{-1}$, where
$B$ is the intensity of the magnetic field, $a$ the grain size, and $n$
the ambient density. If the magnetic field strength scales as $n^{1/2}$
in molecular clouds, as empirically observed (Crutcher~1999), the effect
of the increased density and magnetic field cancel, and one is left with
the dependence of $p/A_V$ on the grain's size (and shape).

Fiege \& Pudritz~(2000) showed that a helical magnetic field geometry
in a cylindrical cloud with uniform grain properties could naturally
produce a depolarization effect, in agreement with the submm polarization
observations of the OMC-3 region of the Orion A filamentary molecular
cloud (Matthews \& Wilson~2000; Matthews, Wilson, \& Fiege~2001).
Padoan et al. (2001) were able to produce a decreasing $p$ vs. $I$
relation in three-dimensional MHD simulations of supersonic and
super-Alfv\'enic turbulence assuming that grains are not aligned above
$A_V\simeq 3$~mag, but the dynamic range of the simulation did not
extend much beyond this value of extinction.  They noticed, however,
that even assuming a uniform grain alignment efficiency, a decrease of
$p$ with $I$ could be reproduced for particular orientations of the core
magnetic field relative to the line of sight, an effect that will be
further explored in the present paper.

It is possible in principle that the observed depolarization may
be accounted for (at least in part) by beam smearing over tangled,
small-scale field structures: Rao et al.~(1998), for example, found that
the decrease in polarization toward the Kleinman-Low nebula previously
seen with single-dish observations was a result of subresolution-scale
variations in the magnetic field that are averaged out by larger beams.
However, given the relatively high levels of polarization detected at
the core peaks, this effect should not be dominant.

We will try in this paper to further analyze the influence of the
magnetic field geometry on the observed depolarization effect, studying
cloud models dominated by a large scale magnetic field. In particular,
we show that the pinching of the magnetic field expected in dense,
self-gravitating molecular cloud cores naturally produces a decrease
of the polarization degree toward the centre of the core, for a large
range of viewing angles. In addition, we find that this geometrical
depolarization effect is further enhanced by a dust temperature gradient
increasing outward, as expected in externally heated starless cores
(Evans et al.~2001, Zucconi et al.~2001; Stamatellos \& Whitworth~2003;
Gon\c {c}alves et al.~2004, hereafter GGW). However, while these effects
can contribute substantially to the observed depolarization, the measured
$p$--$I$ relations are sometimes steeper than our predictions. Thus,
we do not suggest that the field geometry and the dust temperature
distribution are the sole means by which these observations could be
explained: variations in the grain optical and/or geometrical polarization
properties in the densest parts of a cloud are still required to account
for the full range of observed polarization values.

The structure of the paper is the following: in Sect.~2, we describe
the model adopted for magnetized molecular cloud cores and define
the relevant physical and geometrical parameters of the problem; in
Sect.~3, we show synthetic polarization maps computed with these models,
and illustrate the effect of geometrical depolarization as well as its
dependence on inclination, wavelength, and temperature distribution.
In Sect.~4 we summarize our conclusions.

\section{The model}

\begin{figure}[ht]
\resizebox{10cm}{!}{\includegraphics{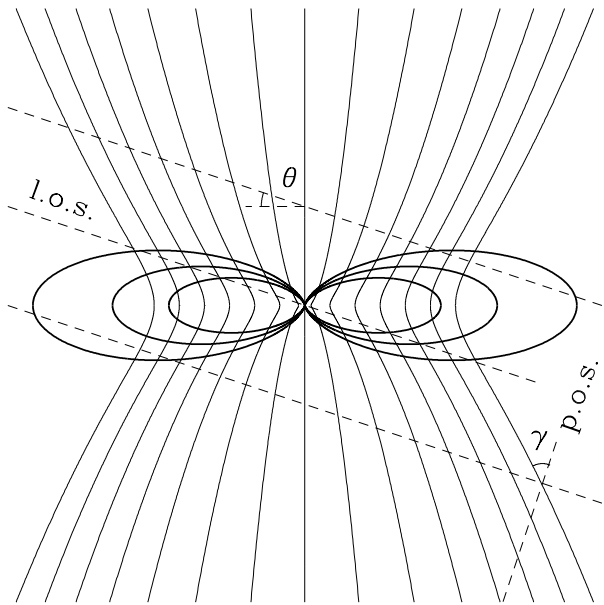}}
\caption{The geometry of the model. The $H_0=1.25$ singular isothermal
toroid ({\it thick curves}: isodensity contours; {\it thin curves}: magnetic
field lines) is observed from a line of sight (l.o.s.) inclined by an angle
$\theta$ with respect to the equatorial plane of the toroid ($\theta=0$: 
edge-on; $\theta=90^\circ$: pole-on). Dust grains are
assumed to be aligned with the local direction of the magnetic field,
making an angle $\gamma$ with the plane of the sky (p.o.s.). Notice that for the
inclination shown in this figure, magnetic field lines are almost in
the plane of the sky for lines of sight intercepting the toroid above
and below the centre, resulting in a relatively higher polarization
degree of the outer parts of the core with respect to the central region.}
\label{model} 
\end{figure}

We model molecular cloud cores as singular isothermal toroids, i.e.
scale-free, axisymmetric equilibrium configurations of an isothermal gas
cloud under the influence of self-gravity, gas pressure and magnetic
forces (Li \& Shu~1996).  These toroids are characterized by one
non-dimensional parameter, $H_0$, representing the fractional amount of
support provided by magnetic forces. We show our polarization results
for three models, with $H_0=0.2$, $H_0=0.5$ and $H_0=1.25$, corresponding
to mass-to-flux ratios (in units of the critical value) $\lambda\approx$
6, 3, and 1.6.  With increasing $H_0$, the density distribution becomes
flatter, and the configuration becomes a thin disk for $H_0\rightarrow
\infty$.  Figure~\ref{model} shows isodensity contours and magnetic field
lines for a singular isothermal toroid with $H_0=1.25$, and illustrates
some geometrical quantities adopted in our analysis.

We further assume that the core is bathed in the average interstellar
radiation field (ISRF) of the solar neighborhood (Black~1994), and use the
results of GGW for the temperature distribution.  The dust temperature
profile and the intensity of the emitted radiation at submillimetre
wavelengths for the $H_0=0.5$ toroid were presented in Sect.~5 and
Fig.~3 of GGW. We note some inconsistencies in our approach:  first, the
core's density distribution was obtained by Li \& Shu~(1996) under the
hypothesis that the gas is isothermal, whereas we explicitly consider
deviations from isothermality at least in the dust component; second,
GGW computed the dust temperature distribution assuming spherical grains,
while in the present work, for the purpose of computing the polarization
of the emitted radiation, we assume that the grains are aspherical and
aligned with the magnetic field.  However, the results of GGW are not
substantially affected by grain shape.

Since we are interested in computing the sub-millimetre thermal emission,
we neglect self-absorption of radiation.  We also assume that the
gas-to-dust ratio and the properties of the dust grains with respect to
absorption and polarizing efficiency are uniform throughout the cloud.
The polarization degree $p$ and polarization
angle $\chi$ (the direction of polarization in the plane of the sky)
are defined in terms of the standard Stokes parameters $Q$, $U$ and $I$,
\be
\label{def}
p = {\sqrt{Q^2+U^2}\over I},
\ee
\be
\label{defchi}
\tan{2\chi} = \frac{U}{Q}.
\ee
Here, we compute $p$ and $\chi$ following a method developed by Lee \&
Draine~(1985), and elaborated by Wardle \& K\"onigl~(1990), Fiege \&
Pudritz~(2000), and Padoan et al.~(2001), but we generalize their
method allowing a dependence of the dust temperature upon position
inside the core.  In this formulation, $p$ and $\chi$ are given by the
simple expressions
\be
\label{Polarization}
p =\alpha \frac{\sqrt{q^2+u^2}}{\Sigma-\alpha\Sigma_2},
\ee
\be
\label{PolarizationAngle}
\tan{2\chi}=\frac{u}{q},
\ee
where $\alpha$ is a nondimensional parameter representing the ``polarizing 
efficiency'' of dust grains, $q$ and $u$ are the ``reduced'' Stokes parameters
\be
q=\int \rho B_\lambda(T_d)\cos{2\psi}\cos^2{\gamma}\; d\ell,
\ee
\be
u=\int \rho B_\lambda(T_d)\sin{2\psi}\cos^2{\gamma}\; d\ell,
\ee
and 
\be
\Sigma = \int \rho B_\lambda(T_d)\; d\ell,
\ee
\be
\Sigma_2 = \int \rho B_\lambda(T_d)
\left({\cos^2\gamma\over 2}-{1\over 3}\right)\; d\ell. 
\ee
These four quantities are integrals along the line of sight $\ell$
of the product of the total density $\rho$ (proportional, under our
hypothesis, to the number density of dust grains), the Planck function
$B_\lambda(T_d)$ (representing the dust emissivity at the dust temperature
$T_d$), and a geometric factor accounting for the orientation of the
magnetic field at each point inside the cloud, characterized by the two
angles $\psi$ and $\gamma$. Specifically, the former is the angle between
a direction in the plane of the sky (e.g.  north) and the component
of ${\bf B}$ in that plane; the latter is the angle between the local
direction of ${\bf B}$ and the plane of the sky (see Fig.~\ref{model}).

Thus, knowledge of the density and temperature distributions $\rho(\ell)$
and $T_d(\ell)$, together with the geometry of the magnetic field in a
cloud core expressed by $\psi(\ell)$ and $\gamma(\ell)$, are sufficient
to completely determine the polarization characteristics of the radiation
emitted by dust grains, if one specifies the nondimensional parameter
$\alpha$ (constant under our assumptions), containing all information
about the absorptions cross sections and the alignment efficiency. The
numerical value of $\alpha$ can be easily fixed observing that the
maximum polarization degree is achieved when ${\bf B}$ is in the plane
of sky. In this case, with $\psi$ constant, eq.~(\ref{Polarization})
gives (Fiege \& Pudritz~2000)
\be
p_{\rm max} = \frac{6\alpha}{6-\alpha},
\label{alpha}
\ee
or 
\be
\alpha=\frac{6p_{\rm max}}{6+p_{\rm max}}\approx p_{\rm max}~~\mbox{if $p_{\rm max}\ll 1$}.
\ee
The choice of $\alpha$ is therefore equivalent to a normalization of
the expected degree of polarization.  The polarization degree and angle
can then be computed fixing the maximum polarization degree, instead of
assuming specific grain properties.  In this work we choose $\alpha=0.15$,
which corresponds, from eq.~(\ref{alpha}), to $p_{\rm max}\approx 15$\%.

\section{Results}

\begin{figure}[ht]
\resizebox{12cm}{!}{\includegraphics[angle=180]{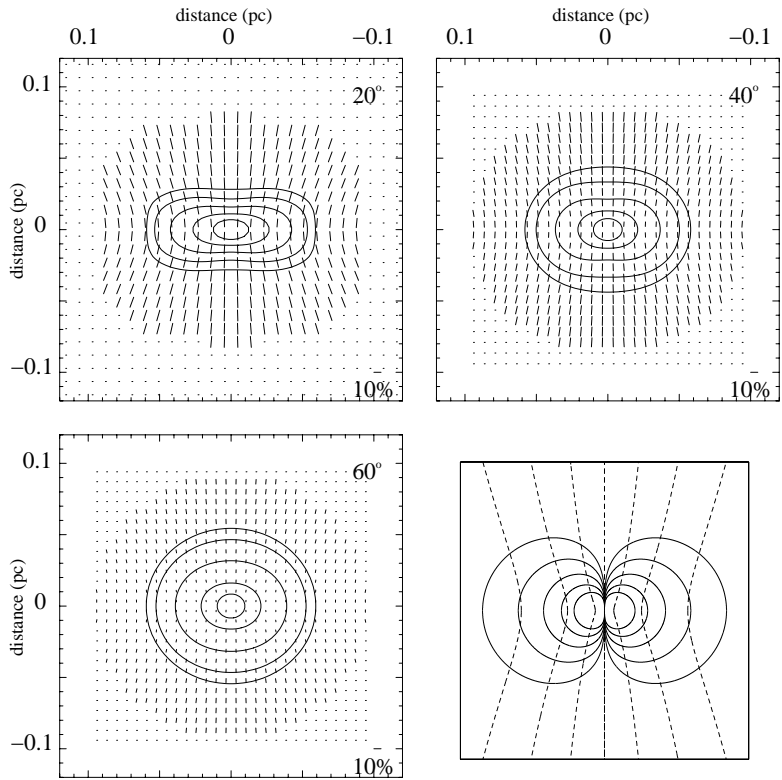}}
\caption{Maps of the dust emission and polarization at
$\lambda=850$~$\mu$m for the $H_0=0.2$ singular isothermal toroid, shown
in the bottom right panel ({\it solid curves}, isodensity contours, {\it
dashed curves}, magnetic field lines). The first three panels are for
inclination with respect to the plane of the sky $\theta=20^\circ$ (top
left panel), $\theta=40^\circ$ (top right panel), and $\theta=60^\circ$
(bottom left panel).  Each vector is proportional to the polarization
degree (see scale on the lower right corner of each panel), and has been
rotated by $90^\circ$ to show the average orientation of the magnetic
field in the plane of the sky.  The intensity is shown by 
contours logarithmically spaced by 0.2 dex starting from 10\%
of the peak value.} 
\label{h02_map}
\end{figure}

\begin{figure}[ht]
\resizebox{12cm}{!}{\includegraphics[angle=180]{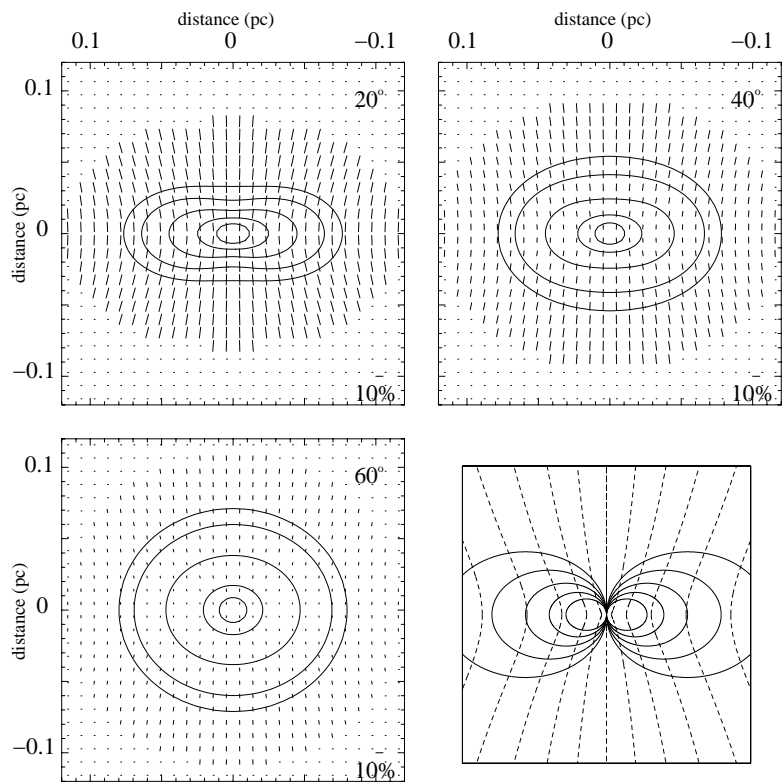}}
\caption{Same as Figure~\ref{h02_map} for the $H_0=0.5$ 
singular isothermal toroid.}
\label{h05_map}
\end{figure}

\begin{figure}[ht]
\resizebox{12cm}{!}{\includegraphics[angle=180]{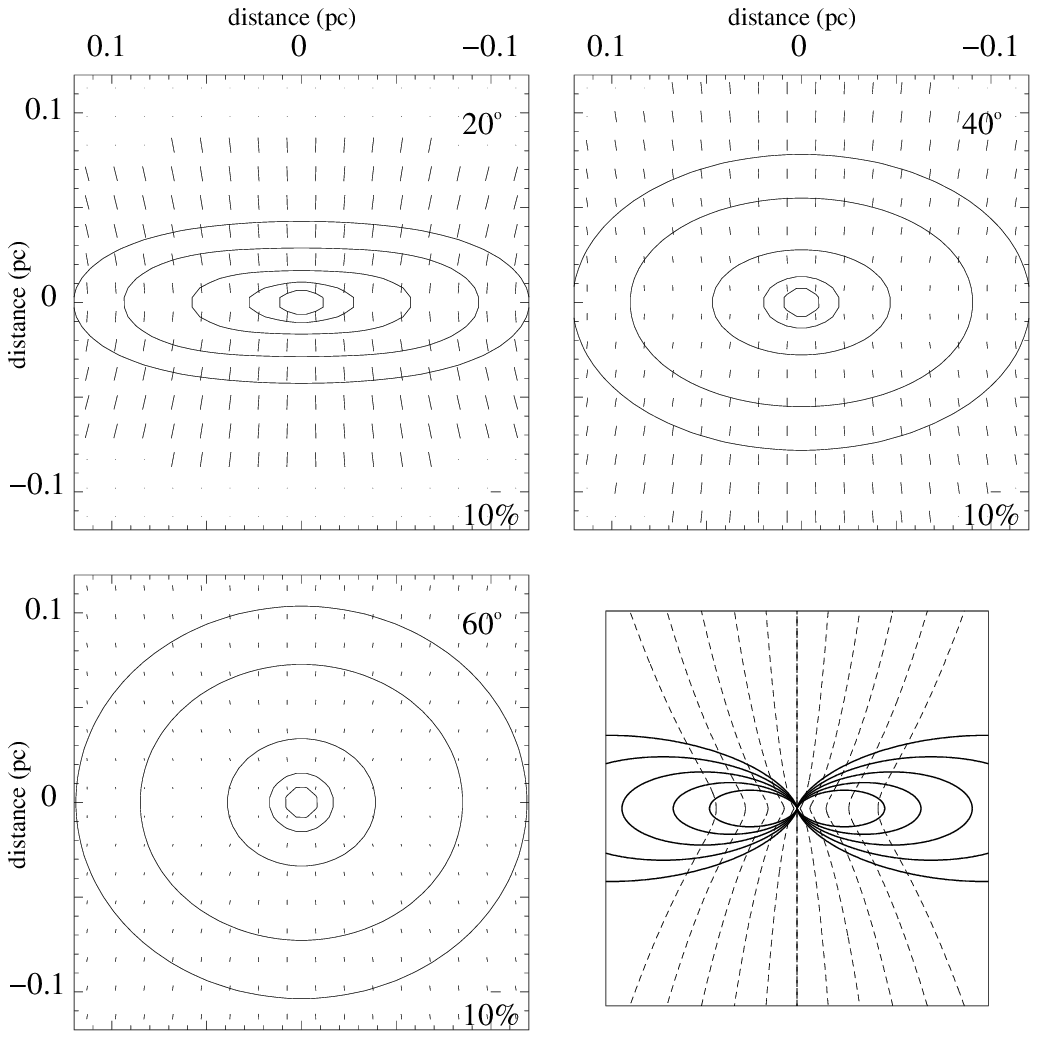}}
\caption{Same as Figure~\ref{h02_map} and \ref{h05_map} for the $H_0=1.25$ 
singular isothermal toroid.}
\label{h125_map}
\end{figure}

For a molecular cloud core modeled as described in Sect.~2, with
specified density and temperature distributions, the only free
parameters are $\alpha$ and the inclination angle $\theta$ of the line
of sight with respect to the equatorial plane of the toroid (
$\theta=0$: edge-on; $\theta=90^\circ$: pole-on).  In this section we
show the polarization of the emitted dust radiation for $\alpha=0.15$,
varying the inclination of the toroid ($\theta=20^\circ$, $40^\circ$,
and $60^\circ$), and for toroids with $H_0=0.2$, 0.5 and 1.25.

\subsection{Geometric depolarization at intermediate inclination}

\begin{figure}[ht]
\resizebox{10cm}{!}{\includegraphics{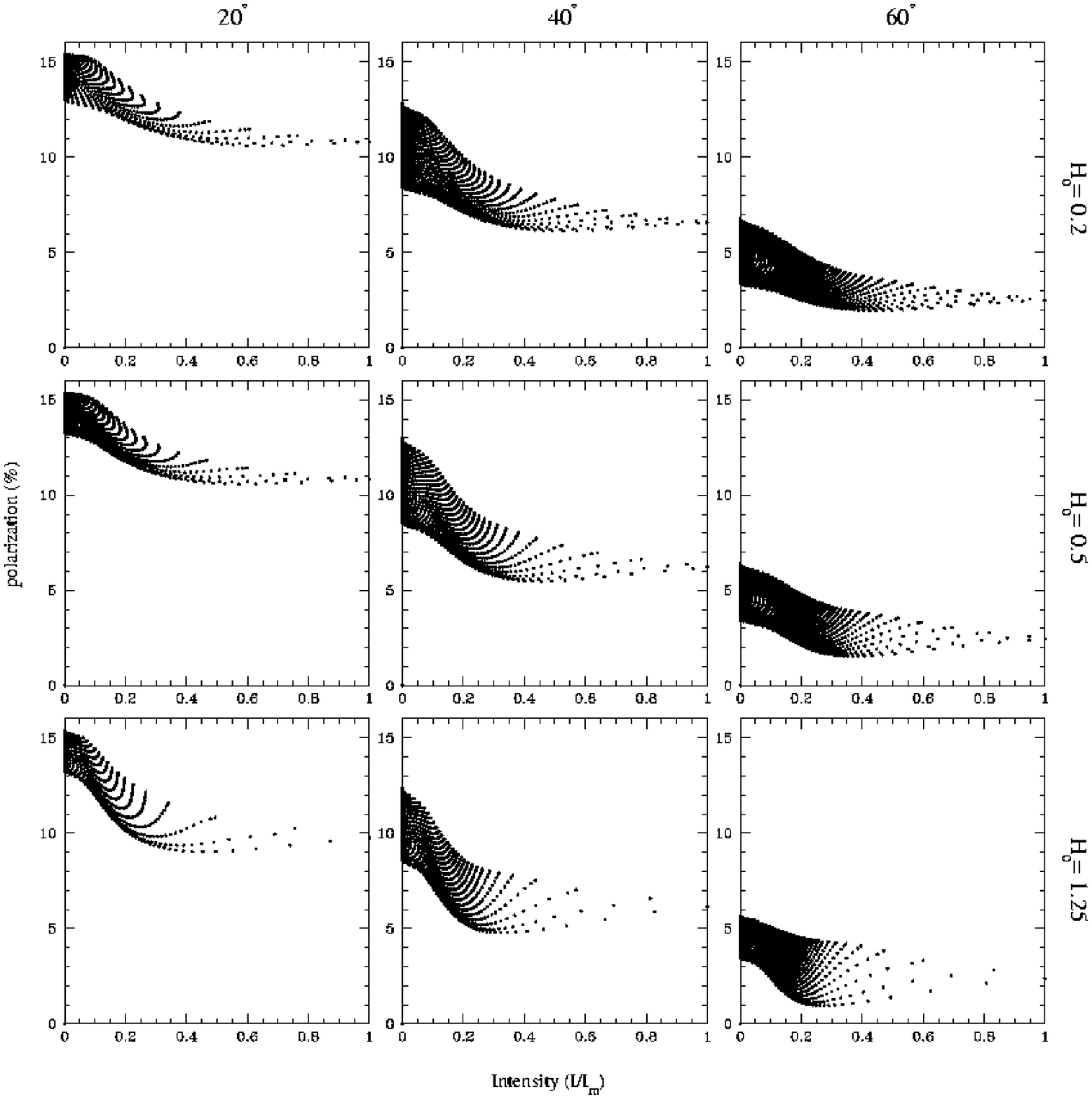}}
\caption{Polarization degree as function of intensity at 850~$\mu$m
(normalized to the peak value $I_{\rm max}$) for the $H_0=0.2$, $0.5$
and $1.25$ toroids for inclinations $\theta=20^\circ$, $40^\circ$
and $60^\circ$. {\bf The multiple tracks visible in each panel are 
an artifact due to the sampling of the polarization degree over a regular 
square grid in the polarization maps.}}
\label{pI}
\end{figure}

In Fig.~\ref{h02_map}, we show the isophotes and polarization vectors
(rotated by $90^\circ$) of the dust emission computed for the $H_0=0.2$
toroid at $\lambda=850$~$\mu$m for $\theta=20^\circ,40^\circ$ and
$60^\circ$. The model results have been convolved with a telescope beam
of FWHS of 12$^{\prime\prime}$, assuming a distance of $150$~pc in all
cases.  Figures~\ref{h05_map} and \ref{h125_map} show the same maps for
the $H_0=0.5$ and 1.25 models, respectively. The polarization vectors
have been rotated by $90^\circ$ to show the approximate average
orientation of the magnetic field in the plane of the sky.

The polarization pattern is clearly symmetric, because of the assumed
axial symmetry of the model. The equatorial pinching of the magnetic
field lines (see Fig.~\ref{model}) has an important effect on the
non-uniformity of the polarization degree across the cloud, resulting
in a significant decrease of $p$ toward the central regions.  In fact,
for a hourglass magnetic field configuration observed at intermediate
inclinations ($\theta\approx 30^\circ$-- $40^\circ$), the largest
component of the field in the plane of the sky is found in the outer
parts of the cloud, whereas lines of sight close to the cloud's centre
intercept regions where the bending of field lines is stronger and
therefore the component of the field in the plane of the sky is
relatively weaker (cf. Fig.~\ref{model}). 

The depolarization effect is better illustrated in Fig.~\ref{pI},
showing the degree of polarization $p$ as function of the intensity at
850~$\mu$m (normalized to the peak intensity) for the $H_0=0.2$, $0.5$
and $1.25$ toroids at three inclinations, $\theta=20^\circ$, $40^\circ$
and $60^\circ$.  The minimum level of polarization is usually attained
toward the centre of the core, where the non uniformity of the magnetic
field along the line of sight is larger, and therefore cancellation
effects more important. Over about one order of magnitude increase in
intensity, the polarization degree decreases with roughly a power-law
behavior. It is interesting to notice that even for a moderate
pinching of the field (e.g. for the $H_0=0.2$ toroid), the decrease of
polarization towards the centre is already significant.  

We do not attempt in this paper to model specific objects, but we
stress the qualitative similarity between our theoretical $p$--$I$
diagrams shown in Fig.~\ref{pI} and the $p$--$I$ relations observed in
dense cores (Fig.~\ref{observations}).  In some cases, however, like in
the dark cloud L183 (Crutcher et al.~2004) or L1544 (Ward-Thomson et
al.~2000), the observed $p$--$I$ relation is too steep to be explained
only on the basis of the field morphology and inclination.  A full
model for the polarization of the emitted radiation probably requires
additional mechanisms (such as an increase in size and sphericity of
dust grains near the core centre) to reduce the polarizing efficiency
$\alpha$ at high values of density or extinction.

\subsection{Dependence of depolarization on aspect ratio}

\begin{figure}[ht]
\resizebox{8cm}{!}{\includegraphics{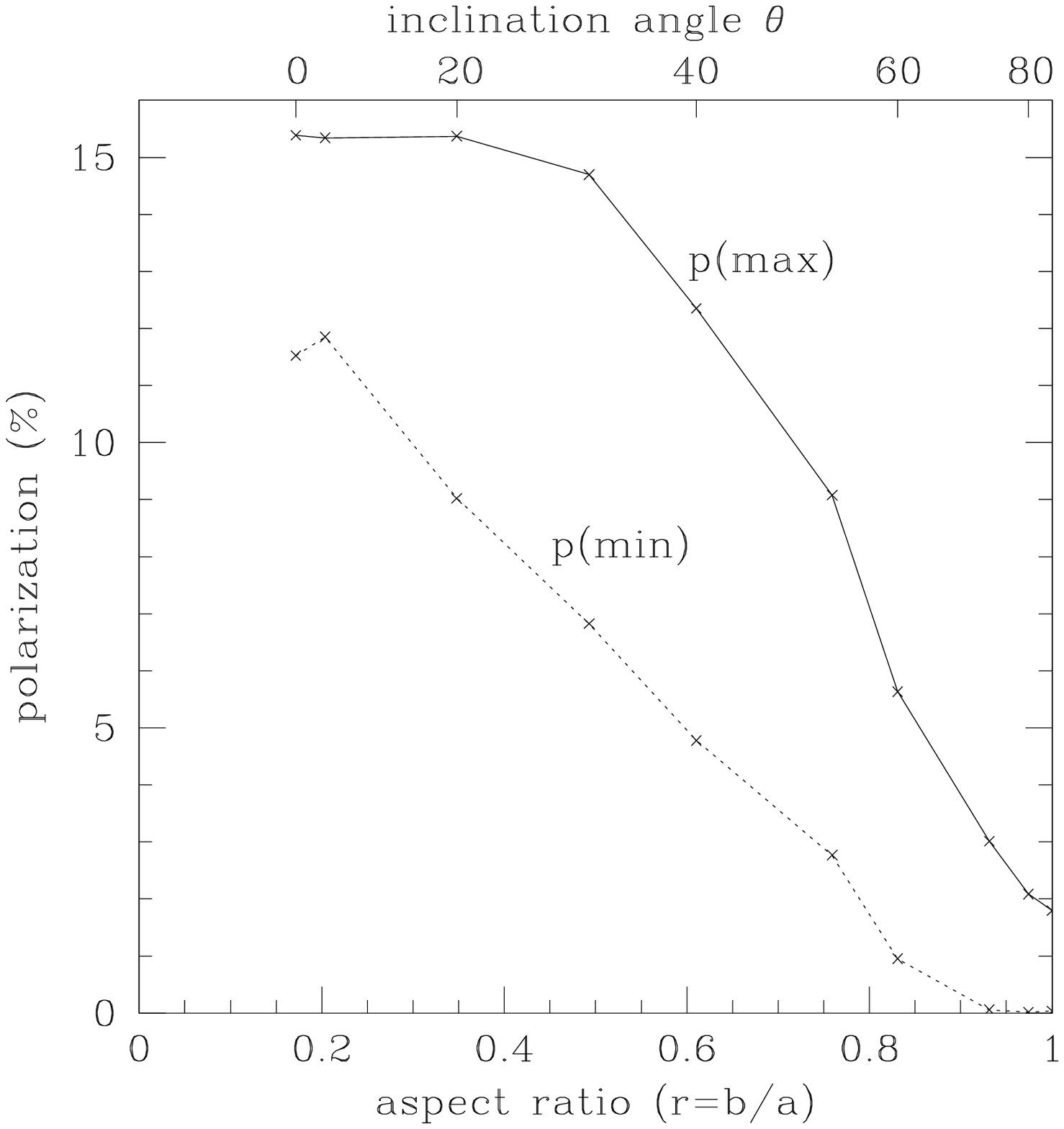}}
\caption{Variation of maximum ({\it solid line}\/) and minimum 
({\it dotted line}\/) polarization degree 
as function of the isophotal aspect ratio $r=b/a$ at $\lambda=850$~$\mu$m
for the $H_0=1.25$ toroid. The upper scale shows the inclination angle 
$\theta$ (in degrees) of the line of sight with respect to the equatorial plane of 
the toroid.}
\label{minmax}
\end{figure}

The maximum and minimum value of the polarization degree, $p_{\rm
max}$ and $p_{\rm min}$, respectively, depend on the inclination angle
$\theta$ of the core with respect to the line of sight. Both $p_{\rm
max}$ and $p_{\rm min}$ decrease when the inclination of the toroid
changes from edge-on ($\theta=0$) to pole-on ($\theta=90^\circ$), where
$p_{\rm min}$ reaches zero. Correspondingly, the isophotal aspect ratio
$r=b/a$ increases from a minimum value, that depends on the adopted model,
to unity. For the three models considered in this paper, the minimum 
value of the aspect ratio is $r_{\rm min}\approx 0.6$ for the $H_0=0.2$
toroid, $r_{\rm min}\approx 0.4$ for the $H_0=0.5$ toroid, and $r_{\rm
min}\approx 0.2$ for the $H_0=1.25$ toroid.

In Fig.~\ref{minmax} we show the maximum and minimum values of
polarization as function of the aspect ratio $r$ for the $H_0=1.25$
toroid at $\lambda=850$~$\mu$m. The inverse correlation shown by
the Figure between polarization degree and core aspect ratio is a
characteristic signature of magnetic field configurations dominated by
a poloidal component, and can be compared to observations to test the
relative importance of poloidal vs. toroidal magnetic field components
in dense clouds. In this sense, it is interesting to notice that of the
three cloud cores represented in Fig.~1, the one showing the smallest
range of variation (and absolute values) of $p$ (B1-c), is also the
one characterized by the largest aspect ratio ($r\approx 1$, compared
to $r \approx 0.5$ for L183 and L1544). A statistical analysis of the
available observations may reveal the presence of a correlation between
the range of polarization degree and the aspect ratio of cloud cores.
As also shown by Fig.~\ref{minmax}, the maximum depolarization effect,
measured by $p_{\rm max}- p_{\rm min}$, is obtained when $r\approx 0.5$
($\theta\approx 30$--$40^\circ$), and corresponds to a polarization
reduction of a factor $\sim 2$, from $p_{\rm max}\approx 15$\% to
$p_{\rm min}\approx 7$\%.  We notice that a detectable decrease in the
polarization degree, say $p_{\rm max}-p_{\rm min}>5$\%, occurs for
inclinations ranging from $\sim 20^\circ$ to $\sim 65^\circ$. For
random orientations of the toroids, this interval corresponds to about
50\% of all possible cases.

\subsection{Dependence of depolarization on wavelength}

\begin{figure}[ht]
\resizebox{10cm}{!}{\includegraphics[angle=-90]{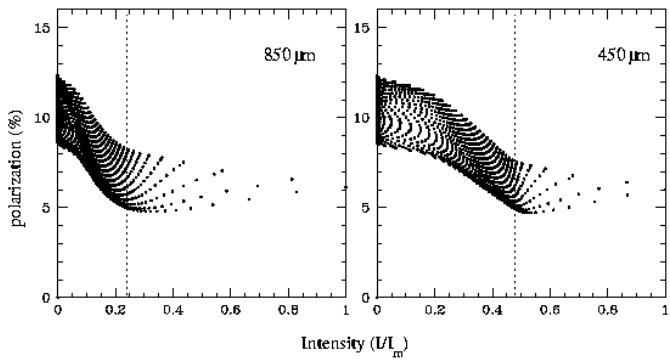}}
\caption{Polarization degree as function of intensity at 850~$\mu$m
({\it left panel}\/) and $450$~$\mu$m ({\it right panel}\/),
normalized to the peak value $I_{\rm max}$, for the $H_0=1.25$ toroid
at $\theta=40^\circ$. Notice the different extent of the ``polarization
hole'', here identified with the region with low and uniform values of
$p$ on the right side of the {\it dashed lines}: within $\sim 80$~\% and $\sim
50$~\% of the peak intensity at 850 and $450$~$\mu$m, respectively.)}
\label{pI_lambda}
\end{figure}

In Fig.~\ref{pI_lambda}, we compare polarization-intensity diagrams
at $\lambda=850$~$\mu$m and $\lambda=450$~$\mu$m, obtained for the
$H_0=1.25$ toroid at $\theta=40^\circ$. As shown by the figure, the
minimum and maximum value of polarization remain the same at these two
wavelengths, but their dependence on intensity is different, the decrease
of $p$ with $I$ being steeper at the longer wavelength. This effect can be
easily understood: since the dust emission at $\lambda=450$~$\mu$m is less
concentrated than at $\lambda=850$~$\mu$m (see e.g. GGW), a given value
of $I/I_{\rm max}$ corresponds to a larger distance from the centre
of the core (the intensity peak) at the shorter wavelength. Therefore the
``polarization hole'' appears restricted to higher values of 
$I/I_{\rm max}$ at $\lambda=450$~$\mu$m.

\subsection{Effects of a non isothermal dust temperature distribution}

\begin{figure}[ht]
\resizebox{10cm}{!}{\includegraphics{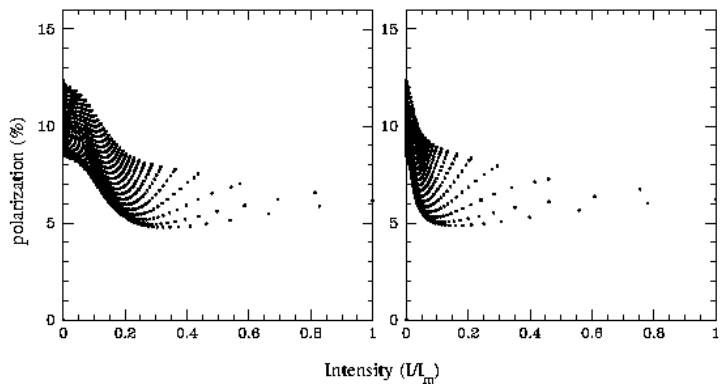}}
\caption{Polarization degree as function of intensity at 850~$\mu$m
(normalized to the peak value $I_{\rm max}$) for the $H_0=1.25$ toroid
at $\theta=40^\circ$, in the case of external heating by the ISRF ({\it
left panel}\/) and assuming an isothermal dust temperature distribution
({\it right panel}\/).}
\label{pI_isot}
\end{figure}

Fig.~\ref{pI_isot} shows the importance of computing the dust
temperature distribution resulting from the heating of the external
ISRF instead of assuming an uniform dust temperature through the
cloud.  In fact, the outwardly increase of the dust temperature from
$\sim 8$~K to $\sim 15$~K for an externally heated cloud (see GGW)
enhances, in the integration along the line of sight, the contribution
to the Stokes parameters of the external layers of the cloud, where, as
we have seen, the component of the magnetic field in the plane of the
sky is generally larger than near the core's centre. Thus, an outward
increasing temperature gradient contributes to the observability of the
depolarization effect.  In fact, given the current sensitivity of
polarimeters at submillimetre wavelengths, that allow measurements of
$p$ only for $I\ga 0.2 I_{\rm max}$ (see Fig.~\ref{observations}), the
value of $p$ would be quite uniform through the sampled region if the
cloud were isothermal (see Fig.~\ref{pI_isot}).  Thus, the decrease of
polarization shown by our model is actually the result of the
combination of pinched magnetic field hourglass {\it and} a dust
temperature gradient increasing outward.

\subsection{Effects of turbulence}

Finally, we remark that the predicted $p$--$I$ diagrams of
condensations formed in simulations of turbulent flows (Padoan et al.
2001) differ significantly from our Fig.~\ref{pI}, as they show a large
population of low-$p$ and low-$I$ data points, which, in general, is
not observed (cf. Fig.~\ref{observations}).  Thus, at least in
principle, $p$--$I$ diagrams offer a way to distinguish between
different explanations of the observed behavior of polarization in
cloud cores, especially given the increased sensitivity expected from 
the next generation of detectors and polarimeters.

In addition, our results show that a large-scale magnetic field with
moderate equatorial pinching can produce a significant deviation
of polarization angles from the direction of the cloud's minor axis, up
to about $\pm 15^\circ$ (see Fig.~3, 4 and 5). For example, in the case
of the $H_0=0.2$ toroid observed at an inclination $\theta=20^\circ$
(Fig.~2), the distribution of polarization angles over the whole map is
peaked on the direction of the cloud's minor axis (a consequence of the
assumed axial symmetry of the model), but the dispersion in
polarization angles around the mean direction is $\sigma_\chi \approx
10^\circ$.  This casts doubt on the use of the Chandrasekhar-Fermi
formula to estimate the magnetic field strength in molecular cloud
cores, as this formula assumes that observed deviations of polarization
angles from a given direction (of the order of $\sigma_\chi\approx
10^\circ$--$15^\circ$ in starless cores, see e.g. Crutcher et al. 2004)
are solely due to the presence of a turbulent (or better, ``wavy'')
component of the field.

\section{Conclusions}

In this paper we have computed the polarization of the radiation emitted
by dust grains in molecular cloud cores represented as magnetically
supported equilibrium configurations. To this end, we have adopted the
magnetostatic models of Li \& Shu~(1996) and the radiative transfer method
developed by GGW. The Stokes parameters (and therefore the polarization
degree) have been computed at two wavelengths ($\lambda=850$~$\mu$m
and $\lambda=450$~$\mu$m) and for various inclinations of the cloud
with respect to the line of sight, assuming that the dust grains are
elongated and aligned with the large-scale magnetic field.

The main result of this paper is the demonstration that a significant
depolarization effect, with characteristics very similar to those
observed in actual cloud cores, can arise only because of geometrical
effects, if the large-scale magnetic field has the equatorially pinched
morphology predicted by magnetically dominated models.  However, we do
not claim that the field geometry is the sole means by which such an
effect could be produced. There may also be contributions due to grain
growth in the densest parts of a cloud, and turbulence.  We also note
that the assumption of axial symmetry of the density and field
distribution implies that the model polarization vectors (after a
$90^\circ$ rotation), are symmetrically distributed with respect to the
cloud's minor axis, at variance with observational evidence for some
cores (see e.g. Ward-Thompson et al.~2000). A toroidal component of the
magnetic field, ignored in the present analysis, can in principle
account for the observed misalignment between the core apparent
elongation and the average polarization position angle (see e.g.
Vall\'ee, Greaves \& Fiege~2003).

\acknowledgements
We thank Richard Crutcher for interesting discussions and for providing 
polarization data for starless cores. JG acknowledges support from
the scholarship SFRH/BD/6108/2001 awarded by the Funda\c c\~ao para a
Ci\^encia e Tecnologia (Portugal).

\bibliography{}

\bibliographystyle{aa}

\end{document}